\documentstyle[amssymb,epsf]{mn_mod}
\def\ulapskip#1#2{\vbox to0pt{\parindent=0mm\vss\hskip#1 #2}}
\def\dot#1{\leavevmode\setbox0\hbox{h}\dimen0\ht0\advance\dimen0-0.75ex
   \rlap{\raise1.40\dimen0\ulapskip{0.39em}{\Large$\cdot$}}#1}
\newcommand{\dm}{\relax\ifmmode{\dot{M}}\else{$\dot{M}$}\fi}
\newcommand{\teff}{\relax\ifmmode{T_{\rm eff}}\else{$T_{\rm eff}$}\fi}
\newcommand{\tstar}{\relax\ifmmode{T_{\ast}}\else{$T_{\ast}$}\fi}

\newcommand{\pta}{\relax\ifmmode{P_{\rm a}}\else{$P_{\rm a}$}\fi}
\newcommand{\ptb}{\relax\ifmmode{P_{\rm b}}\else{$P_{\rm b}$}\fi}
\newcommand{\md}{\mbox{\rm d}}
\newcommand{\me}{\mbox{\rm e}}
\newcommand{\p}{{\mathchoice
  {\setbox0=\hbox{$\displaystyle'$}\kern0.25\wd0\lower0.28\ht0\box0}
  {\setbox0=\hbox{$\textstyle'$}\kern0.25\wd0\lower0.28\ht0\box0}
  {\setbox0=\hbox{$\scriptstyle'$}\lower0.28\ht0\box0}
  {\setbox0=\hbox{$\scriptscriptstyle'$}\lower0.28\ht0\box0}}}
\newcommand{\pp}{\relax\ifmmode{^\prime}\else{$^\prime$}\fi}
\newcommand{\binom}[2]{\left(\!\!\begin{array}{c} #1 \\ #2 \end{array}\!\!\right)}
\newcommand{\fc}{f_{\rm c}}
\newcommand{\rs}{r_{\rm s}}
\newcommand{\ri}{r_{\rm i}}
\newcommand{\rii}{\rs}
\newcommand{\Phd}{\Phi_{\rm d}}
\newcommand{\Phb}{\Phi_{\rm b}}
\newcommand{\dia}{\Theta_{\rm d}}

\newcommand{\kms}{\relax\ifmmode{\rm km\,s^{-1}}\else{km\,s$^{-1}$}\fi}

\newcommand{\mic}{\relax\ifmmode{\mu{\rm m}}\else{$\mu$m}\fi}
\newcommand{\zm}{\relax\ifmmode{\rm M_\odot}\else{M$_\odot$}\fi}
\newcommand{\zl}{\relax\ifmmode{\rm L_\odot}\else{L$_\odot$}\fi}
\newcommand{\zs}{\relax\ifmmode{\rm R_\odot}\else{R$_\odot$}\fi}
\newcommand{\zmy}{\relax\ifmmode{\rm M_\odot\,yr^{-1}}\else{M$_\odot$\,yr$^{-1}$}\fi}
\newcommand{\ky}{\relax\ifmmode{\rm K\,yr^{-1}}\else{K\,yr$^{-1}$}\fi}
\newcommand{\wsm}{\relax\ifmmode{\rm W\,m^{-2}}\else{W\,m$^{-2}$}\fi}
\newcommand{\escs}{\relax\ifmmode{\rm erg\,cm^{-2}\,s^{-1}}\else{erg\,cm$^{-2}$\,s$^{-1}$}\fi}
\newcommand{\rscs}{\relax\ifmmode{\rm Ryd\,cm^{-2}\,s^{-1}}\else{Ryd\,cm$^{-2}$\,s$^{-1}$}\fi}
\newcommand{\fhb}{\relax\ifmmode{F\rm (H\beta)}\else{$F$(H$\beta$)}\fi}
\newcommand{\lhb}{\relax\ifmmode{L\rm (H\beta)}\else{$L$(H$\beta$)}\fi}
\newcommand{\jb}{{\rm\sl B}}
\newcommand{\jv}{{\rm\sl V}}

\newcommand{\av}{\relax\ifmmode{A_{\rm V}}\else{$A_{\rm V}$}\fi}
\newcommand{\ebv}{\relax\ifmmode{E(\jb-\jv)}\else{$E(\jb-\jv)$}\fi}

\newcommand{\w}{\relax\ifmmode{\lambda}\else{$\lambda$}\fi}
\newcommand{\lya}{\relax\ifmmode{\rm Ly\alpha}\else{\rm Ly$\alpha$}\fi}
\newcommand{\ha}{\relax\ifmmode{\rm H\alpha}\else{\rm H$\alpha$}\fi}
\newcommand{\hb}{\relax\ifmmode{\rm H\beta}\else{\rm H$\beta$}\fi}
\newcommand{\hg}{\relax\ifmmode{\rm H\gamma}\else{\rm H$\gamma$}\fi}
\newcommand{\hd}{\relax\ifmmode{\rm H\delta}\else{\rm H$\delta$}\fi}
\newcommand{\he}{\relax\ifmmode{\rm H\epsilon}\else{\rm H$\epsilon$}\fi}
\newcommand{\bra}{\relax\ifmmode{\rm Br\alpha}\else{\rm Br$\alpha$}\fi}
\newcommand{\brg}{\relax\ifmmode{\rm Br\gamma}\else{\rm Br$\gamma$}\fi}
\newcommand{\pfa}{\relax\ifmmode{\rm Pf\alpha}\else{\rm Pf$\alpha$}\fi}

\newcommand{\al}[2]{#1\,{\sc #2}}
\newcommand{\odd}{\rlap{\rm o}}
\newcommand{\od}[3]{\smash{\relax\ifmmode{^{#1}{\rm #2}^{\odd}_{#3}}\else{$^{#1}$#2$^{\odd}_{#3}$}\fi}}
\newcommand{\ev}[3]{\smash{\relax\ifmmode{^{#1}{\rm #2}_{#3}}\else{$^{#1}$#2$_{#3}$}\fi}}

\newcommand{\fwhm}{{\sc fwhm}}

\newcommand{\str}{Str\"omgren}

\newcommand{\eq}{Eq.}
\newcommand{\eqs}{Eqs.}
\newcommand{\req}[1]{\eq~(\ref{#1})}
\newcommand{\Req}[1]{Eq.~(\ref{#1})}
\newcommand{\reqs}[1]{\eqs~(\ref{#1})}
\newcommand{\reqb}[1]{\eq~\ref{#1}}

\newcommand{\sct}{Section}

\newcommand{\fig}{Fig.}

\newcommand{\x}[1]{\hspace*{#1mm}}
\newdimen\mbtindent
\def\mbtlist#1{\setbox0=\hbox{#1\,--\,}\mbtindent=\wd0\vspace{\baselineskip}
\bgroup\hsize=\columnwidth\leftskip=\mbtindent%
\rightskip=0pt\parindent=-\mbtindent\lineskiplimit=-10pt\parskip=0pt}
\def\endmbtlist{\egroup}
\newcommand{\mbt}[2]{\leavevmode\hbox to\mbtindent{#1\hfil--\,}#2\par}

\newcounter{lemma}
\def\lem{\refstepcounter{lemma}\paragraph*{Lemma \thelemma}\bgroup\it}
\def\endlem{\egroup}

\def\bbbn{{\mathbb{N}}}
\def\int{\intop}
\def\harvarditem#1#2#3{\bibitem[#2]{#3}}
\def\bibcode#1{}
\begin{document}

\pubyear{1999}

\onecolumn

\title[Measuring angular diameters of extended sources -- Theory]
{Measuring angular diameters of extended sources\\I. Theory}
\author[P.A.M. van Hoof]{P.A.M. van Hoof\\
Kapteyn Astronomical Institute, P.O. Box 800, 9700 AV Groningen,
The Netherlands\\
University of Kentucky, Department of Physics and Astronomy,
177 CP Building, Lexington, KY 40506--0055, USA}
\date{printed: \today}
\maketitle

\begin{abstract}
When measuring diameters of partially resolved sources like planetary
nebulae, \al{H}{ii} regions or galaxies, often a technique
called gaussian deconvolution is used.
This technique yields a gaussian diameter which subsequently has to be
multiplied with a conversion factor to obtain the true
angular diameter of the source.
This conversion factor is a function of the Full Width at Half Maximum
(\fwhm) of the beam (in the case of radio observations) or the point spread
function (in the case of optical observations). It also depends on the
intrinsic surface brightness distribution of the source.

In this paper the theory behind this technique will be studied.  This study
will be restricted to circularly symmetric geometries and beams.  First an
implicit equation will be derived, from which the conversion factor for a
given surface brightness distribution and beam size can be solved.  Explicit
expressions for the conversion factor will be derived from this equation which
are valid in cases where the beam size is larger than the intrinsic size of
the source.  A more detailed discussion will be given for two simple
geometries: a circular constant surface brightness disk and a spherical
constant emissivity shell with arbitrary inner radius.  The theory is
subsequently used to construct a new technique for determining the
\fwhm\ of an arbitrary observed surface brightness distribution.

Usually the \fwhm\ of the source and beam are measured using gaussian fits,
but second moments can also be used.  The alternative use of second moments in
this context is studied here for the first time and it is found that in this
case the conversion factor has a different value which is {\em independent} of
the beam size.  In the limit for infinitely large beam sizes, the values of
the conversion factors for both techniques are equal.

The application of the theory discussed in this paper to actual observations
will be discussed in a forthcoming paper.  This will include a comparison
between optical and radio observations.
\end{abstract}


\section{Introduction}

The accurate measurement of angular diameters is a long standing problem.
This problem is pertinent to the study of planetary nebulae, \al{H}{ii}
regions, galaxies and other extended sources. Nevertheless, only few papers
dedicated to this problem can be found in the literature: Mezger \& Henderson
\cite{c3:mezger}, Panagia \& Walmsley \cite{c3:pana}, Bedding \& Zijlstra
\cite{c3:bed:zijl}, Schneider \& Buckley \cite{c3:schneider}, and 
Wellman et al. \cite{c3:wellman}.

Several methods are in general use to determine angular diameters.  One
method determines the angular diameter of the source, based on the Full Width
at Half Maximum (\fwhm) of a two-dimensional gaussian fitted to the observed
surface brightness distribution (also called profile in this
paper) in a least-squares sense.  This method is usually called gaussian
deconvolution and will be explained in more detail below.  The second method
is basically identical to the first, except that it determines the \fwhm\
using the second moment of the profile instead of a gaussian fit. To
discriminate it from the first method, it will be called second moment
deconvolution.  Both methods have the disadvantage
that they yield a result that has no well-defined physical meaning. Hence, a
conversion factor is needed to translate the result into something
meaningful. In nebular research this usually is the \str\ radius.  This
conversion factor depends on the method being used, the intrinsic surface
brightness profile of the source and the resolution of the observation.

The theory to determine the conversion factor is derived in this paper in
several stages. First, in \sct~\ref{comput} some basic assumptions and
definitions will be given. Next, in \sct~\ref{form} the full analytical theory
for gaussian deconvolution will be given. In \sct~\ref{alter} this theory will
be used to construct an alternative algorithm to determine the \fwhm\ of an
observed surface brightness distribution. In \sct~\ref{sec:mom} the theory
will be given for second moment deconvolution.  In \sct~\ref{concl:iv} the
main conclusions will be presented.  In Appendix~\ref{notes} some additional
proofs will be presented and finally in Appendix~\ref{symbols:iv} the most
important symbols that have been used will be defined. The appendices are
only available in electronic form.

\section{Assumptions and definitions}
\label{comput}

The methods discussed in this article can be applied to observations at any
wavelength. More in particular, they are valid for optical, infrared and radio
observations. The resolution of these observations is usually characterized by
the size of the beam profile for radio data or by the size of the point spread
function for optical and infrared data.  Throughout the paper the term `beam'
will be used and it will be implicitly understood that it can also mean `point
spread function' where appropriate.  It will be assumed that the beam can be
approximated by a gaussian.  In this paper the intrinsic surface brightness
profile will be defined as the profile that would be observed with a perfect
instrument (i.e. an instrument with infinite resolving power).  For simplicity
it will be assumed throughout this paper that both the surface brightness
distribution of the nebula and the beam are circularly symmetric.  This is a
rather severe restriction; nebulae rarely are circular, and also for radio
observations the beam usually is elliptical.  However, this simplified case
already yields interesting results which can be applied to actual data.

As was already remarked, a conversion factor is needed to translate the \fwhm\
diameter yielded by the gaussian or second moment deconvolution method into a
\str\ diameter.  In this paper the \str\ radius of the nebula will be denoted
by $\rs$.  In the rest of this paper it will be assumed that the true diameter
of the nebula is $\dia = 2\rs$.  The measured \fwhm\ of the nebula will be
denoted by $\Phi$ and the \fwhm\ of the beam by $\Phb$.  Throughout the paper
the deconvolved \fwhm\ diameter $\Phd$ will be used, which is defined by
\begin{equation}
  \Phd = \sqrt{\Phi^2 - \Phb^2}.
\label{phd:def}
\end{equation}
This quantity is also commonly called the gaussian diameter.  It should not
be confused with the \fwhm\ of the deconvolved or intrinsic profile, which in
general will be different.  The conversion factor to obtain the true
angular diameter from the deconvolved \fwhm\ can now be defined as
\begin{equation}
  \dia = \gamma \Phd \x{3} \Rightarrow \x{3} \gamma = \dia/\Phd = 2\rs/\Phd.
\label{convfac}
\end{equation}
This conversion factor is a function of the resolution of the observation, or
to be more precise, of the ratio of the observed source diameter and the beam
size.  Hence an independent parameter $\beta$ is chosen, which is defined as
\begin{equation}
  \beta = \Phd/\Phb.
\label{beta:def}
\end{equation}
In the following sections more details will be given of the techniques that
have been used to calculate the conversion factors, both for the gaussian
and the second moment deconvolution method.

\section{Determining diameters using gaussian fits}
\label{form}

In this section the conversion factor will be studied for the case where the
\fwhm\ of the observed surface brightness profile is determined using gaussian
fits. An implicit equation will be derived from which the value of
$\gamma(\beta)$ can be solved for arbitrary $\beta$. This equation will also
be used to derive the first terms of a Taylor series expansion of
$\gamma(\beta)$ near $\beta = 0$.  This implicit equation is derived in three
stages. First it will be shown how to derive the width of a gaussian fitted to
an arbitrary profile in \sct~\ref{gauss:fit}.  Next an expression for a
surface brightness profile convolved with a gaussian of arbitrary size will be
derived in \sct~\ref{conv:prf}.  Finally in \sct~\ref{width} these results
will be combined to determine the conversion factor.  In this process various
lemma's will be used which can be found in Appendix~\ref{notes}.
\begin{list}{}{\leftmargin=2.0em\itemsep=1.0ex}
\item[{\em Note 1:}]
in the remainder of this paper all functions are implicitly assumed to be
circularly symmetric.
\item[{\em Note 2:}]
in the remainder of this paper constraints have to be imposed
on the surface brightness profile $f(r)$.
When $c_n$ is defined as
\begin{equation}
  c_n = 2\pi\!\int_0^\infty f(r) r^{n+1} \md r,
\label{std:moment}
\end{equation}
the following constraint can be formulated:
\[ \hbox{(1)} \x{5} 0 \leq f(r) \leq F \x{3} \hbox{for all } r\in[\,0,\infty),
   \x{3} \hbox{such that $c_0$ exists and } c_0 > 0. \]
This condition is sufficient for the theory derived in \sct~\ref{gauss:fit}.
For the theory in \sct~\ref{conv:prf} (up to \reqb{conv:func}) an
additional constraint has to be imposed
\[
  \hbox{(2a)} \x{4} c_{2n} \hbox{ exists for all } n\in\bbbn_0,
  \x{3} \hbox{with } c_n^{1/n} = o(n)\,(n\rightarrow\infty).
\]
It should be noted that the fact that $c_n$ exists automatically implies
that all $c_i$ exist with $i \leq n$, as is shown in Lemma~\ref{l:cn}.
In the remainder of \sct~\ref{conv:prf} and all subsequent sections,
this constraint needs to be restricted to
\[
  \hbox{(2b)} \x{4} c_{2n} \hbox{ exists for all } n\in\bbbn_0,
  \x{3} \hbox{with } c_n^{1/n} = o(n^\frac{1}{2})\,(n\rightarrow\infty).
\]
It is also worth noting that all profiles which fulfill the following condition
\[ \hbox{(2c)} \x{4} f(r) = 0 \x{3} \hbox{for all } r > R_o, \]
(where $R_o$ is some arbitrary outer radius) and which additionally fulfill
condition~(1), automatically fulfill condition~(2b). This is proven in
Lemma~\ref{cn:lim}. Since for all observational
data conditions~(1) and (2c) are automatically fulfilled, the theory
presented in this paper is valid for all observed profiles.
This neglects the fact that noise might cause some pixels to have negative
flux values. In general this poses no problem however, as will be discussed in
a forthcoming paper (van Hoof 1999). One should
also note that for a perfect gaussian, condition (2b) is {\em not} fulfilled.
\item[{\em Note 3:}]
in the remainder of this paper it is assumed that $\rs = 1$.
\end{list}

\subsection{Fitting a gaussian to a surface brightness profile}
\label{gauss:fit}

In this section the formulae to approximate a surface brightness profile $f(r)$
with a two dimensional gaussian in a least-squares sense will be derived.  The
gaussian will be written as
\[ g(r;a,p) = a\,\me^{-pr^{2}} \hspace{3mm}\hbox{with}\ a>0,\ p>0. \]
In order to determine a least squares fit to the surface brightness profile,
the integral over the quadratic residuals $\chi^2$ needs to be minimized. 
Hence the following two equations need to be solved:
\begin{equation}
   \frac{\partial\chi^2}{\partial p} =
   \frac{\partial}{\partial p} \int_0^{2\pi} \int_0^\infty
   \left[\,f(r) - g(r;a,p)\,\right]^2 r \md r \md \varphi = 0,
\label{eqnp}
\end{equation}
and
\begin{equation}
   \frac{\partial\chi^2}{\partial a} =
   \frac{\partial}{\partial a} \int_0^{2\pi} \int_0^\infty
   \left[\,f(r) - g(r;a,p)\,\right]^{2} r \md r \md \varphi = 0.
\label{eqna}
\end{equation}
\Req{eqnp} will be evaluated first. After integration over $\varphi$ and
reversal of the order of differentiation and integration one gets
\[ -4\pi \! \int_0^\infty \left[\,f(r) - g(r;a,p)\,\right]
   \frac{\partial g(r;a,p)}{\partial p} r \md r = 0. \]
One can easily prove that this is allowed using Lemma~\ref{l:exi} and
\ref{l:iunc}, provided condition~(1) is fulfilled.  Since
\[ \frac{\partial g(r;a,p)}{\partial p} = -ar^2 \me^{-pr^2} \]
one finds, after division by the constants,
\[ \int_0^\infty \left[\,f(r) - a\me^{-pr^2}\,\right]\,r^2
   \me^{-pr^{2}} r \md r = 0, \]
and thus
\begin{equation}
  \int_0^\infty  f(r)\,r^3\,\me^{-pr^2} \md r =
  a  \int_0^\infty  r^3\,\me^{-2pr^2} \md r = \frac{a}{8p^2}.
\label{resp}
\end{equation}
For \req{eqna} one finds after a similar derivation that
\begin{equation}
   \int_0^\infty f(r)\,r\,\me^{-pr^2} \md r = \frac{a}{4p}.
\label{resa}
\end{equation}
Hence
\[   \int_0^\infty f(r)\,r\,\me^{-pr^2} \md r  =
 2p \int_0^\infty f(r)\,r^3\,\me^{-pr^2} \md r, \]
and thus the width of the gaussian fitted to the surface brightness profile can
be found by solving \req{impl:r}.
\begin{equation}
  I(p) = \int_0^\infty f(r)\,(1 - 2pr^2)\,r\,\me^{-pr^2}\md r = 0.
\label{impl:r}
\end{equation}
If the substitution $s = pr^{2}$ is used, one can write
\begin{equation}
  I(p) = \frac{1}{2p} \int_0^\infty f\!\left(\sqrt{s/p}\,\,\right)\,
  (1 - 2s)\,\me^{-s}\md s=0.
\label{impl}
\end{equation}
Once the width has been determined, the height of the gaussian can be solved
from either \req{resp} or \req{resa}.
Using \req{impl:r} one can easily see that
\[ \lim_{p\rightarrow0} I(p) = \frac{c_0}{2\pi} \x{1} > \x{1} 0. \]
Substituting a Taylor series expansion of $f$ in \req{impl} one can prove that
for $p\rightarrow\infty$ the following is true
\[ I(p) = -\,f^{(n)}(0)\,\,\frac{(n+1)\,\Gamma\!\left( \frac{n}{2}+1 \right)}{2\,n!}\,
   p^{-\frac{n}{2}-1} + O\left( p^{-\frac{n}{2}-\frac{3}{2}}
   \right) \x{1} < \x{1} 0, \]
where $f^{(n)}(0)$ with $n\in\bbbn_0$ denotes the first non-zero
derivative of $f$. Its value must be positive due to constraint (1).
Because $I$ is a continuous function in $p$, its limiting behavior assures that
at least one solution of \req{impl} must exist.  Since
\[ \frac{\partial^2\chi^2}{\partial a^2} = \frac{\pi}{p} \x{1} > \x{1} 0, \]
this solution can either be a saddle point or a minimum depending on the sign of
\begin{equation}
   \frac{\partial^2\chi^2}{\partial p^2} = -2\pi\frac{a}{p}\int_0^\infty
   \frac{\md f}{\md r}\,r^4\,\me^{-pr^2} \md r.
\label{chi:deriv}
\end{equation}
From this equation it is clear that for monotonically decreasing $f(r)$ all
solutions of \req{impl} must be minima. Given the fact that $\chi^2$ and its
derivatives are continuous functions in both $a$ and $p$, this implies that
only one minimum can exist and thus the solution of \req{impl} must be unique.
In general however, this is not the case.

\subsubsection*{The constant surface brightness disk}

\Req{impl} usually gives rise to implicit equations which can only be
solved numerically. For a constant surface brightness disk the intrinsic
surface brightness profile is very simple and a reasonably simple expression
for the \fwhm\ of the fitted gaussian and the conversion factor can de derived.
\begin{equation}
   f(r) = \left\{ \begin{array}{ll}
                 f_0 & \mbox{\hskip 2mm $r \leq \rs$} \vspace{1mm} \\
                 0 & \mbox{\hskip 2mm $r > \rs$}
                 \end{array}
          \right.
\label{diskprof}
\end{equation}
Without loss of generality it can be assumed that $f_0 = 1/\pi$ and $\rs = 1$.
One can easily see that this profile satisfies constraints (1) and (2c).
Substituting \req{diskprof} in \req{chi:deriv} one can also prove that a
unique fit must exist. Hence, to
determine the width of the gaussian fitted to the unconvolved profile, one may
substitute \req{diskprof} in \req{impl}:
\[
  \frac{1}{2p} \int_0^{p} f_0 \, (1 - 2s) \, \me^{-s} \md s = 0
  \x{3} \Rightarrow \x{3}
  \frac{1}{2p\pi} \left[ \, (2p+1)\,\me^{-p} - 1 \right] = 0.
\]
Multiplying this equation by $2p\pi$ and taking the natural logarithm yields
the following implicit equation for $p$
\begin{equation}
  p = \ln(2p+1) = 1.2564312\ldots \x{3} \Rightarrow \x{3}
  \Phi(p) = 1.4855024\ldots \x{3} \Rightarrow \x{3}
  \gamma(\infty) = 1.3463458\ldots
\label{diskinf}
\end{equation}

\subsection{An expression for the convolved surface brightness profile}
\label{conv:prf}

In this section an expression for an intrinsic surface brightness profile
convolved with a gaussian of arbitrary size will be derived. This way a
relation between the intrinsic and the observed profile will be found.
The intrinsic (unconvolved) surface brightness profile will be written as
$f(r)$ and the convolved profile as $\fc(r)$. Hence
\begin{equation}
 \fc(x,y) = \int_{-\infty}^\infty \int_{-\infty}^\infty
    f(x\p,y\p)\,g(x-x\p,y-y\p;a,p)\,\md x\p \md y\p =
    a \int_{-\infty}^\infty \int_{-\infty}^\infty
    f(x\p,y\p)\,\me^{-p(x-x\p)^{2}} \me^{-p(y-y\p)^{2}} \md x\p \md y\p.
\label{conv}
\end{equation}
The fact that the gaussian representing the beam should be normalized to 1
implies $a = p/\pi$.  In order to evaluate \req{conv} the exponential
functions will be replaced by their Taylor series in the following way:
\begin{equation}
  \me^{-p(x-x\p)^{2}} = \me^{-px^2} \me^{-px\p^2} \me^{2pxx\p} =
  \me^{-px^2} \me^{-px\p^2} \sum_{n=0}^\infty \frac{(2pxx\p)^n}{n!}.
\label{gtayl}
\end{equation}
Since it can be proved easily that the series expansion in \req{gtayl} is
absolutely and uniformly convergent with an infinite radius of convergence
both in $x$ and $x\p$, it is allowed to substitute \req{gtayl} in
\req{conv}. Hence
\[ \fc(x,y) = a \int_{-\infty}^\infty \int_{-\infty}^\infty f(x\p,y\p)
  \, \me^{-px^2} \me^{-px\p^2} \me^{-py^2} \me^{-py\p^2}
  \left[ \sum_{n=0}^\infty \frac{(2pxx\p)^n}{n!} \right]
  \left[ \sum_{n=0}^\infty \frac{(2pyy\p)^n}{n!} \right] \md x\p \md y\p = \]
\[ \phantom{\fc(x,y)} = a\,\me^{-p(x^2+y^2)}
  \int_{-\infty}^\infty \int_{-\infty}^\infty f(x\p,y\p)
  \, \me^{-p(x\p^2+y\p^2)} \sum_{n=0}^\infty \, (2p)^n \sum_{i=0}^n
  \frac{(xx\p)^{n-i}(yy\p)^i}{(n-i)!\,i!} \md x\p \md y\p. \]
When $x\p$, $y\p$ are changed to polar coordinates, this equation can be
rewritten as
\begin{equation}
  \fc(x,y) = a\,\me^{-p(x^2+y^2)} \int_0^\infty \int_0^{2\pi} f(r\p)
  \, \me^{-pr\p^2} \sum_{n=0}^\infty \, (2p)^n \sum_{i=0}^n
  \frac{x^{n-i}y^i}{(n-i)!\,i!}\,r\p^n\cos^{n-i}\!\varphi\p\,\sin^i\!\varphi\p\,
   r\p \, \md \varphi\p \, \md r\p.
\label{lem:int}
\end{equation}
When conditions~(1) and (2a) are fulfilled, the order of the integration and
summation may be reversed (see Lemma~\ref{l:sunc1}) and one finds
\begin{equation}
  \fc(x,y) = a\,\me^{-p(x^2+y^2)} \int_0^\infty f(r\p) \, \me^{-pr\p^2}
  \sum_{n=0}^\infty \, (2p)^n r\p^{n+1} \sum_{i=0}^n
  \frac{x^{n-i}y^i}{(n-i)!\,i!} \int_0^{2\pi} \cos^{n-i}\!\varphi\p\,
  \sin^i\!\varphi\p\, \md \varphi\p \, \md r\p.
\label{lab1}
\end{equation}
The inner integral is well known and yields a non-zero result only when both
$n-i$ and $i$ are even. In this case the result is
\begin{equation}
  \int_0^{2\pi} \cos^{2n-2i}\!\varphi \, \sin^{2i}\!\varphi \, \md\varphi=
  2\pi \, \frac{(2n-2i)!\,(2i)!}{2^{2n}\,n!\,(n-i)!\,i!}.
\label{lab2}
\end{equation}
When \req{lab2} is substituted in \req{lab1} and all the odd terms in
$n$ and $i$ are omitted one finds
\[ \fc(x,y) = a\,\me^{-p(x^2+y^2)} \int_0^\infty f(r\p)\,\me^{-pr\p^2}
  \sum_{n=0}^\infty \, (2p)^{2n} r\p^{2n+1} \sum_{i=0}^n
  \frac{x^{2n-2i}y^{2i}}{(2n-2i)!\,(2i)!} \, 2\pi \,
  \frac{(2n-2i)!\,(2i)!}{2^{2n}\,n!\,(n-i)!\,i!} \, \md r\p = \]
\[ \phantom{\fc(x,y)} = 2\pi\,a\,\me^{-p(x^2+y^2)}\int_0^\infty f(r\p)\,
  \me^{-pr\p^2} \sum_{n=0}^\infty \, (2p)^{2n} r\p^{2n+1} \frac{1}{2^{2n}\,n!}
  \sum_{i=0}^n \frac{x^{2n-2i}y^{2i}}{(n-i)!\,i!} \, \md r\p = \]
\[ \phantom{\fc(x,y)} = 2\pi\,a\,\me^{-p(x^2+y^2)}\int_0^\infty f(r\p)\,
  \me^{-pr\p^2} \sum_{n=0}^\infty\frac{p^{2n}}{n!^2}\,r\p^{2n+1} \sum_{i=0}^n
  \binom{n}{i} x^{2(n-i)}y^{2i}
  \, \md r\p = \]
\[ \phantom{\fc(x,y)} = 2\pi\,a\,\me^{-p(x^2+y^2)}\int_0^\infty f(r\p)\,
  \me^{-pr\p^2} \sum_{n=0}^\infty \frac{p^{2n}}{n!^2}\,r\p^{2n+1} (x^2+y^2)^n
  \, \md r\p. \]
When $x$, $y$ are changed to polar coordinates and the order of
integration and summation is reversed one gets (Lemma~\ref{l:sunc1} proves that
this is allowed)
\begin{equation}
  \fc(r) = 2\pi\,a\,\me^{-pr^2} \sum_{n=0}^\infty \frac{p^{2n}}{n!^2}\,r^{2n}
  \int_0^\infty f(r\p) \, r\p^{2n+1} \me^{-pr\p^2} \md r\p.
\label{lab3}
\end{equation}
When the generalized $n$-th moment of $f(r)$ is defined as
\begin{equation}
  c_n(p) = 2\pi \! \int_0^\infty f(r) \, r^{n+1} \me^{-pr^2} \md r,
\label{gen:moment}
\end{equation}
\req{lab3} can be rewritten to the following expression for the convolved
surface brightness profile
\begin{equation}
 \fc(r)=a\,\me^{-pr^2}\sum_{n=0}^\infty\frac{c_{2n}(p)\,p^{2n}}{n!^2}\,r^{2n}.
\label{conv:func}
\end{equation}
In Lemma~\ref{l:cnp} it will be proven that this expression is absolutely
and uniformly convergent for all $p,r\in[0,\infty)$,
provided that conditions~(1) and (2a) are fulfilled.
An alternative formulation for \req{conv:func} can be derived which will be
used further on.
When the exponential in \req{gen:moment} is expanded in a Taylor series in $p$
(which can easily be shown to have an infinite radius of convergence) one finds
\begin{equation}
  c_n(p) = 2\pi \! \int_0^\infty f(r) \, r^{n+1} \sum_{k=0}^\infty\,
  (-1)^k \frac{(pr^2)^k}{k!} \md r.
\label{lem:int2}
\end{equation}
When conditions~(1) and (2b) are fulfilled, the order of integration and
summation may be reversed (Lemma~\ref{l:sunc2}), hence
\[ c_n(p) = \sum_{k=0}^\infty\,(-1)^k \, \frac{p^k}{k!} \, 2\pi
  \! \int_0^\infty f(r)\, r^{n+2k+1} \md r. \]
When \req{std:moment} is used, one can write
\begin{equation}
  c_n(p) = \sum_{k=0}^\infty\,(-1)^k\, \frac{p^k}{k!}\, c_{n+2k}.
\label{tayl:moment}
\end{equation}
In Lemma~\ref{l:sunc5} it will be proven that this series is absolutely
and uniformly convergent for all $p\in[0,\infty)$,
provided condition~(2b) is fulfilled.
Hence one can substitute \req{tayl:moment} in \req{conv:func}
\begin{equation}
  \fc(r) = a\,\me^{-pr^2}\sum_{n=0}^\infty\frac{p^{2n}r^{2n}}{n!^2}
  \sum_{k=0}^\infty\,(-1)^k \,\frac{p^k}{k!} \,c_{2n+2k}.
\label{lem:int5}
\end{equation}
Lemma~\ref{l:cnp} and Lemma~\ref{l:sunc5} imply that this series is absolutely
and uniformly convergent for all $p\in[0,\infty)$ and $r\in[0,\infty)$,
provided conditions~(1) and (2b) are fulfilled.
Hence the summation may be reordered in any way.
The following transformation will be used: $n\rightarrow k\pp$,
$k\rightarrow(n\pp - k\pp)$, which yields (after dropping the primes)
\begin{equation}
  \fc(r) = a\,\me^{-pr^2}\sum_{n=0}^\infty c_{2n}\,p^n \sum_{k=0}^n
  (-1)^{n-k} \, \frac{(pr^2)^k}{(n-k)!\,k!^2}.
\label{conv:func2}
\end{equation}
Closer inspection reveals that the inner summation is closely related to
the Laguerre polynomials $L_n$. Hence one can write
\begin{equation}
  \fc(r) = a\,\me^{-pr^2}\sum_{n=0}^\infty (-1)^n \frac{c_{2n}\,p^n}{n!}
  L_n(pr^2).
\label{conv:laguerre}
\end{equation}

\subsection{The width of the convolved surface brightness profile}
\label{width}

Now an expression for the convolved profile has been found, the next task is to
determine the width $p\p$ of a gaussian fitted to this profile. This is
equivalent to measuring the \fwhm\ of an observed profile. The result can be
found by substituting \req{conv:func2} in \req{impl}. This is allowed since
\req{conv:func2} is absolutely and uniformly convergent for all
$p\in[0,\infty)$ and $r\in[0,\infty)$. The result is
\[
   \frac{1}{2p} \int_0^\infty (1 - 2s) \, \me^{-s} \fc\left(\sqrt{s/p\p}
   \right) \md s = 0 \x{3} \Rightarrow \x{3}
   \frac{1}{2p} \int_0^\infty (1 - 2s) \, \me^{-s} a\,\me^{-s\frac{p}{p\p}}
   \sum_{n=0}^\infty c_{2n}p^n \sum_{k=0}^n (-1)^{n-k}\,\frac{s^k}
   {(n-k)!\,k!^2} \left( \frac{p}{p\p} \right)^k \! \md s = 0.
\]
When one abbreviates $\zeta = \frac{p}{p\p} + 1$ and multiplies with $2p/a$,
one can write
\begin{equation}
   \int_0^\infty (1 - 2s) \, \me^{-\zeta s} \sum_{n=0}^\infty\,
   c_{2n}p^n \sum_{k=0}^n (-1)^{n-k}\,\frac{(\zeta - 1)^k}{(n-k)!\,k!^2} s^k
   \md s = 0.
\label{lem:int4}
\end{equation}
Provided condition~(2b) is fulfilled, the order of summation and integration
may be reversed (see Lemma~\ref{l:sunc4}), hence
\[
   \sum_{n=0}^\infty\,
   c_{2n}p^n \sum_{k=0}^n (-1)^{n-k}\,\frac{(\zeta - 1)^k}{(n-k)!\,k!^2}
   \int_0^\infty (1 - 2s) \, s^k \, \me^{-\zeta s} \md s = 0.
\]
It is well known that
\[ \int_0^\infty x^n\,\me^{-px} \md x = \frac{n!}{p^{n+1}}. \]
Hence
\[
   \sum_{n=0}^\infty\,
   c_{2n}p^n \sum_{k=0}^n (-1)^{n-k}\,\frac{(\zeta - 1)^k}{(n-k)!\,k!^2}
   \left[ \frac{k!}{\zeta^{k+1}} - 2 \frac{(k+1)!}{\zeta^{k+2}} \right] = 0 
   \x{3} \Rightarrow \x{3}
   \sum_{n=0}^\infty c_{2n} p^n \sum_{k=0}^n\,(-1)^{n-k}\,
   \frac{[\,\zeta-2(k+1)\,]}{(n-k)!\,k!}\frac{(\zeta - 1)^k}
   {\zeta^{k+2}} = 0 \x{3} \Rightarrow
\]
\begin{equation}
   \sum_{n=0}^\infty \frac{c_{2n} p^n}{n!} \sum_{k=0}^n\,(-1)^{n-k}
   \binom{n}{k}
   \left[\,\zeta-2(k+1)\,\right]\frac{(\zeta - 1)^k}{\zeta^{k+2}} = 0.
\label{lab4}
\end{equation}
Usually the width of a gaussian is given as the Full Width at Half Maximum
(\fwhm) value $\Phi$. One can easily see that the \fwhm\ for the beam $\Phb$ is
given by
\begin{equation}
  \Phb(p) = 2\sqrt{\frac{\ln 2}{p}}.
\label{phi1}
\end{equation}
The deconvolved \fwhm\ $\Phd$ of a fitted gaussian can be defined as
\begin{equation}
  \Phd(p\p,p) = \sqrt{\Phi^{2}(p\p) - \Phb^{2}(p)}.
\label{phi2}
\end{equation}
Substituting \req{phi1} in \req{phi2} gives
\begin{equation}
  \Phd(p\p,p) = 2\sqrt{\frac{\ln2}{p\p} - \frac{\ln2}{p}} =
   2\sqrt{\frac{\ln2}{p}\left(\frac{p}{p\p} - 1\right)} =
   2\sqrt{\frac{\ln2}{p}(\zeta - 2)}.
\label{phi3}
\end{equation}
When one recalls the definition for $\beta$ in \req{beta:def} a
relation between $\zeta$ and $\beta$ can be obtained
\begin{equation}
  \beta = \frac{\Phd(p\p,p)}{\Phb(p)} = 2\sqrt{\frac{\ln2}{p}(\zeta - 2)}
  \left/ 2\sqrt{\frac{\ln2}{p}} \right. = \sqrt{\zeta-2} \x{3}
  \Rightarrow \x{3} \zeta = \beta^2 + 2.
\label{zeta:beta}
\end{equation}
When \req{convfac} and \req{phi3} are combined, a relation between
$p$ and $\gamma$ can be found
\begin{equation}
  \gamma(\beta) = \frac{2}{\Phd(p\p,p)} =
  \sqrt{\frac{p(\beta)}{(\zeta - 2)\ln2}} = \sqrt{\frac{p(\beta)}{\beta^2\ln2}}
  \x{3} \Rightarrow \x{3} p(\beta) = \gamma^2(\beta)\,\beta^2\ln2.
\label{p:gamma2}
\end{equation}
When \reqs{zeta:beta} and (\ref{p:gamma2}) are substituted in \req{lab4} one
finds
\[ \sum_{n=0}^\infty \frac{c_{2n}\,\gamma^{2n}(\beta)\,\beta^{2n} \ln^n2}{n!}
   \sum_{k=0}^n\,(-1)^{n-k}\binom{n}{k}
   (\beta^2-2k) \frac{(\beta^2 + 1)^k}{(\beta^2+2)^{k+2}} = 0. \]
If it is assumed that $\beta>0$ this equation can be multiplied by
$(\beta^2+2)^2/\beta^2$. This yields the following implicit equation for
$\gamma(\beta)$
\begin{equation}
   \sum_{n=0}^\infty \frac{c_{2n} \gamma^{2n}(\beta) \ln^n2}{n!}
   \sum_{k=0}^n\,(-1)^{n-k}\binom{n}{k}
   (\beta^2-2k) \, \beta^{2n-2} \left( \frac{\beta^2+1}{\beta^2+2}
   \right)^k = 0.
\label{gen:eqn}
\end{equation}
Now one can determine $\gamma(0)$ by taking the limit
$\beta\rightarrow0$ of the left-hand side of \req{gen:eqn}. This yields
\begin{equation}
   c_0 - c_2\gamma^2(0)\ln2 = 0 \x{3}\Rightarrow\x{3} \gamma(0) =
   \sqrt{\frac{c_0}{c_2\ln2}}.
\label{gamma0}
\end{equation}
This yields the following expression for the deconvolved \fwhm.
\begin{equation}
  \Phd(0,0) = 2\sqrt{\frac{c_2\ln2}{c_0}}.
\label{deconv}
\end{equation}
So it can be seen that the deconvolved \fwhm\ of the source in the limit
for very large beams
is fully determined by two simple integrals over the intrinsic surface
brightness profile.  When \req{deconv} is compared with \req{sec:def}
one can see that this limiting value simply is the \fwhm\ derived
from the second moment of the unconvolved profile.
In \sct~\ref{sec:mom} it will be shown that this is also the deconvolved
\fwhm\ derived from the second moment of a profile convolved
with a beam of arbitrary width.

\Req{gen:eqn} can also be used to determine
a value for $\gamma(\beta)$ for arbitrary $\beta$. It can easily be
solved numerically using a Newton-Raphson scheme.
It can be shown that the Taylor series of $\gamma(\beta)$ contains only
even powers of $\beta$. Hence one may write
\[ \gamma(\beta) = \sum_{l=0}^\infty g^{\phantom{()}}_{2l}\beta^{2l}. \]
It should be noted that in general this expansion need not have an infinite
radius of convergence, hence this expression is only valid sufficiently close
to $\beta = 0$.
Now the $2n$-th power of this series can be calculated, which will have the same
radius of convergence as the original series
\begin{equation}
  \gamma^{2n}(\beta) = \sum_{l=0}^\infty g^{(2n)}_{2l}\beta^{2l}
  \x{3}\hbox{with}\x{3} g^{(2n)}_0 = g_0^{2n\rlap{\phantom{()}}};
  \x{3} g^{(2n)}_{2l} = \frac{1}{2l\,g^{\phantom{()}}_0}\sum_{k=1}^{l}\,
  \left[\,2k\,(2n+1) - 2l\,\right]\,g^{\phantom{()}}_{2k}g^{(2n)}_{2l-2k}.
\label{gn:ser}
\end{equation}
When \req{gn:ser} is substituted in \req{gen:eqn} one finds
\begin{equation}
   \sum_{n=0}^\infty \frac{c_{2n} \ln^n2}{n!} \sum_{l=0}^\infty g^{(2n)}_{2l}
   \sum_{k=0}^n\,(-1)^{n-k}\binom{n}{k}
   (\beta^2-2k) \, \beta^{2n+2l-2} \left( \frac{\beta^2+1}{\beta^2+2}
   \right)^k = 0.
\label{gen:eqn2}
\end{equation}
Taking the 2nd derivative, the 4th derivative, etc\ldots\ to $\beta$ of
\req{gen:eqn2} and subsequently taking the limit $\beta\rightarrow0$, new
equations are obtained from which $g_2^{\phantom{|}}$, $g_4^{\phantom{|}}$,
etc\ldots\ can be solved.  Without proof the following result will be given
\begin{equation}
  \gamma(\beta) = \sqrt{\frac{c_0}{c_2\ln2}}\, \left(\, 1 -
  \frac{2c_2^2 - c_0c_4}{4c_2^2}\,\beta^2 +
  \frac{20c_2^4-18c_0c_2^2c_4+7c_0^2c_4^2-2c_0^2c_2c_6}{32c_2^4}\,\beta^4
  \right) +  O(\beta^6)
  \,(\beta\rightarrow0).
\label{final:gamma}
\end{equation}
Again it is noted that this series expansion in general need not have an
infinite radius of convergence, hence it is only valid sufficiently close to
$\beta = 0$.

\subsection{Application of the theory}

\subsubsection{The constant surface brightness disk}

The coefficients $c_{2n}$ for the constant surface brightness disk can be
calculated by substituting \req{diskprof} in \req{std:moment}. This gives
\[ c_{2n} = 2\pi \int_0^1 \frac{1}{\pi} r^{2n+1} \md r =
   \frac{2}{2n+2}. \]
Substituting this result in \req{final:gamma} yields (where additional terms are
given without proof)
\begin{equation}
  \gamma(\beta) = \sqrt{\frac{2}{\ln2}} \left( 1 - \frac{1}{6}\beta^2 +
  \frac{5}{36}\beta^4 - \frac{43}{360}\beta^6 + \frac{1367}{12960}\beta^8 -
  \frac{51817}{544320}\beta^{10}
 + \frac{57119}{653184}\beta^{12} 
  \right) + O(\beta^{14})
  \,(\beta\rightarrow0).
\label{diskzero}
\end{equation}
Inspection of this expression suggests that the radius of convergence $B$
is approximately $B\approx1$.

\subsubsection{The constant volume emissivity shell}

\begin{figure}
\begin{center}
\mbox{\epsfxsize=0.45\textwidth\epsfbox[48 352 468 696]{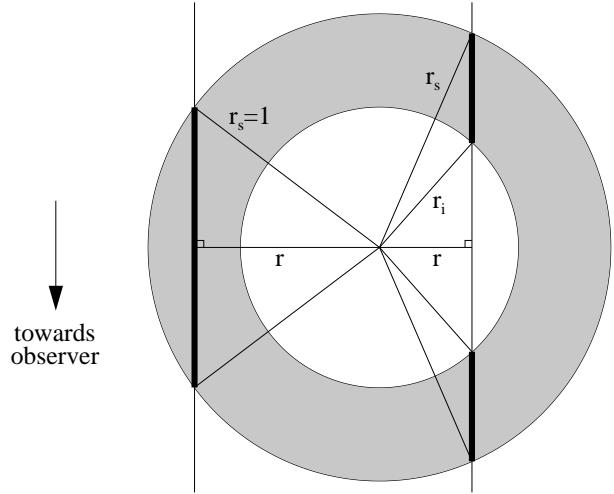}}
\caption
{The geometry for a constant volume emissivity shell is shown.  Two lines of
sight through the shell are also shown. The surface brightness for these lines
of sight is proportional to the length of the thick parts of the lines.}
\label{shell:geo}
\end{center}
\end{figure}

\noindent
The geometry for the constant volume emissivity shell is shown in
\fig~\ref{shell:geo}. Using this figure one can easily understand that the
surface brightness for any given line of sight is proportional to the length of
that part of the line that actually passes through the nebula.  Hence for a
constant emissivity shell the surface brightness profile is given by
\begin{equation}
\vspace{2.5mm}
   f(r) = \left\{ \begin{array}{l@{\x{5}}l}
                 2j_\nu \! \left( \sqrt{\rii^2-r^2} - \sqrt{\ri^2-r^2} \right) &
                 r \! \leq \! \ri \! < \! \rii \\
                 2j_\nu\sqrt{\rii^2-r^2} & \ri \! < \! r \! \leq \! \rii
                 \phantom{\rlap{$\left(\sqrt{\rii^2}\right)$}} \\
                 0 & \ri \! < \! \rii \! < \! r
                 \phantom{\rlap{$\left(\sqrt{\rii^2}\right)$}}
                 \end{array}
          \right.
\vspace{2.5mm}
\label{shellprof}
\end{equation}

\noindent
Without loss of generality it can be assumed that
$j_\nu = \frac{1}{4\pi}$ and $\rii = 1$.
One can easily see that these profiles satisfy conditions~(1) and (2c).
First an analytic expression for $c_{2n}$ will be deduced by substituting
\req{shellprof} in \req{std:moment}
\[ c_{2n} = \left( \int_0^1 r^{2n+1}\sqrt{1-r^2} \md r -
  \int_0^{\ri} r^{2n+1}\sqrt{\ri^2-r^2} \md r \right). \]
These integrals can be solved easily by using the substitution $t = r^2$
and give
\begin{equation}
  c_{2n} = \sum_{k=0}^n\,(-1)^{k+n}\binom{n}{k}\frac{1 - \ri^{2n+3}}{2n+3-2k}.
\label{shelli}
\end{equation}

\noindent
The coefficients in the limiting case for an infinitely thin shell can also be
calculated. After normalization to $c_0 = 1$ one finds
\begin{equation}
  c_{2n} = \sum_{k=0}^n\,(-1)^{k+n} \binom{n}{k} \frac{2n+3}{2n+3-2k}.
\label{shell:lim}
\end{equation}
Substituting \req{shelli} in \req{gamma0} gives
\begin{equation}
  \gamma(0) = \sqrt{\frac{5(1-\ri^3)}{2(1-\ri^5) \ln2}}.
\label{shellphii}
\end{equation}
When \req{shelli} is substituted in \req{final:gamma} the following
expression for the spherical case $\ri$ = 0 is found (again giving
additional terms without proof):
\begin{equation}
  \gamma(\beta) = \sqrt{\frac{5}{2\ln2}} \left( 1 - \frac{1}{7}\beta^2 +
  \frac{5}{42}\beta^4 - \frac{1157}{11319}\beta^6 + 
  \frac{2226013}{24720696}\beta^8
  - \frac{42074119}{519134616}\beta^{10}
  \right) + O(\beta^{12})
  \,(\beta\rightarrow0).
\label{spherezero}
\end{equation}
When \req{shell:lim} is substituted in \req{final:gamma} one finds
for the limiting case $\ri\rightarrow 1$
\begin{equation}
  \gamma(\beta) = \sqrt{\frac{3}{2\ln2}} \left( 1 - \frac{1}{5}\beta^2 +
  \frac{59}{350}\beta^4 - \frac{128}{875}\beta^6 + \frac{70117}{539000}\beta^8
  - \frac{20675393}{175175000}\beta^{10}
  \right) + O(\beta^{12})
  \,(\beta\rightarrow0).
\label{limitzero}
\end{equation}
Inspection of both expressions suggests that the radius of convergence $B$
is approximately $B\approx1$.

\section{An alternative algorithm for determining the FWHM diameter}
\label{alter}

In this section, an expression will be derived that can be used to determine
the \fwhm\ diameter of an arbitrary surface brightness profile. This method
will constitute an alternative algorithm to determine the \fwhm\ diameter
which is fully equivalent to a gaussian fit algorithm. The derivation will
start with \req{gen:eqn}:
\begin{equation}
   \sum_{n=0}^\infty (-1)^n \frac{c_{2n} \, \gamma^{2n}(\beta) \ln^n2}{n!}
   \sum_{k=0}^n\,(-1)^k\binom{n}{k}
   (\beta^2-2k) \, \beta^{2n-2} \left( \frac{\beta^2+1}
   {\beta^2+2}\right)^k = 0.
\label{gen:inf0}
\end{equation}
This expression will be transformed into an expression for the conversion
factor of the intrinsic surface brightness profile by taking the limit
$\beta\rightarrow\infty$. This will be done by substituting $\beta = 1/\epsilon$
and subsequently taking the limit $\epsilon\rightarrow0$.
First the innermost summation (which will be called $S_n$) will be evaluated.
\[
   S_n = \sum_{k=0}^n\,(-1)^k\binom{n}{k}
   \left( \frac{1}{\epsilon^2}-2k \right) \, \epsilon^{2-2n} \left(
   \frac{1+\epsilon^2} {1+2\epsilon^2} \right)^k =
   \sum_{k=0}^n\,(-1)^k\binom{n}{k}
   (1-2k\epsilon^2) \, \epsilon^{-2n} \left[ (1+\epsilon^2)
   \sum_{l=0}^\infty(-1)^l \, 2^l \, \epsilon^{2l}\right]^k =
\]
\begin{equation}
   \phantom{S_n} = \sum_{k=0}^n\,(-1)^k\binom{n}{k}
   (1-2k\epsilon^2) \, \epsilon^{-2n} \left[ 1 +
   \sum_{l=1}^\infty(-1)^l\,2^{l-1}\,\epsilon^{2l}\right]^k
   \equiv \sum_{k=0}^n\,(-1)^k\binom{n}{k}
   (1-2k\epsilon^2) \, \epsilon^{-2n}
   \sum_{l=0}^\infty a_{k,l} \epsilon^{2l}.
\label{sn0}
\end{equation}
The Taylor expansion of $(1 + 2\epsilon^2)^{-1}$ has a radius of convergence of
$\frac{1}{2}\sqrt{2}$, hence this derivation is only valid for $0 \leq
\epsilon < \frac{1}{2}\sqrt{2}$.  The coefficients $a_{k,l}$ have been defined
such that:
\begin{equation}
   \left[ 1 + \sum_{l=1}^\infty(-1)^l\,2^{l-1}\,\epsilon^{2l}\right]^k \equiv
   \sum_{l=0}^\infty a_{k,l} \epsilon^{2l} \x{3} \Rightarrow \x{3}
   a_{k,0} = 1;
   \x{3} a_{k,l} = \frac{1}{l}\sum_{m=1}^{l}[\,m(k+1) - l\,]\,(-1)^m\,2^{m-1}
   \,a_{k,l-m}.
\label{sn4}
\end{equation}
This expression for $a_{k,l}$ will yield a polynomial in $k$ of degree $l$,
which will be written as (see Lemma~\ref{lem:akl}):
\[
   a_{k,l} = \sum_{i=0}^{l} d_{l,i} k^i = (-1)^l\frac{k^l}{l!} + \sum_{i=0}^{l-1}
   d_{l,i} k^i \equiv (-1)^l\frac{k^l}{l!} + b_{k,l} \x{3} {\rm for\ all\ }
   k,l\in\bbbn_0 {\rm \ and\ where\ } 0^0 \equiv 1.
\]
Here $b_{k,l}$ represents the terms in $k$ with powers less than $l$ (if any).
It can easily be shown that the series given by \req{sn4} is absolutely and
uniformly convergent within its radius of convergence $\frac{1}{2}\sqrt{2}$.
Hence it is allowed to substitute the last expression in \req{sn0} and
reorder the summation:
\[
   S_n = \sum_{k=0}^n\,(-1)^k\binom{n}{k}
   (1-2k\epsilon^2) \, \epsilon^{-2n}
   \sum_{l=0}^\infty \left[ (-1)^l\frac{k^l}{l!} + b_{k,l} \right]
   \epsilon^{2l} = 
\]
\[
   \phantom{S_n} = \sum_{k=0}^n\,(-1)^k\binom{n}{k} \, \left[
   \sum_{l=0}^\infty (-1)^l\frac{k^l}{l!} \epsilon^{2l-2n} (1-2k\epsilon^2) +
   \sum_{l=0}^\infty b_{k,l} \, \epsilon^{2l-2n} (1-2k\epsilon^2) \right].
\]
When the limit $\epsilon\rightarrow0$ is taken, terms with $l > n$ will tend
to zero, hence the following is true:
\[
   \lim_{\epsilon\rightarrow0} S_n = \lim_{\epsilon\rightarrow0} \sum_{k=0}^n\,(-1)^k
   \binom{n}{k} \, \left[
   \sum_{l=0}^n (-1)^l\frac{k^l}{l!} \epsilon^{2l-2n} (1-2k\epsilon^2) +
   \sum_{l=0}^n b_{k,l} \, \epsilon^{2l-2n} (1-2k\epsilon^2) \right].
\]
When the order of the summation is reversed, this gives:
\[
   \lim_{\epsilon\rightarrow0} S_n = \lim_{\epsilon\rightarrow0} \sum_{l=0}^n \sum_{k=0}^n \,
   (-1)^k \binom{n}{k} \,
   (-1)^l\frac{k^l}{l!} \epsilon^{2l-2n} (1-2k\epsilon^2) \ + \ 
   \sum_{l=0}^n \sum_{k=0}^n \,
   (-1)^k \binom{n}{k} \,
   b_{k,l} \, \epsilon^{2l-2n} (1-2k\epsilon^2) = 
\]
\[
   \phantom{\lim_{\epsilon\rightarrow0} S_n} = \lim_{\epsilon\rightarrow0}
   \sum_{l=0}^n (-1)^l \frac{\epsilon^{2l-2n}}{l!} \sum_{k=0}^n \,
   (-1)^k \binom{n}{k} k^l \ + \ 
   \sum_{l=0}^{n-1} (-1)^{l+1} \frac{2\epsilon^{2l-2n+2}}{l!} \sum_{k=0}^n \,
   (-1)^k \binom{n}{k} k^{l+1} \ +
\]
\begin{equation}
   \phantom{\lim_{\epsilon\rightarrow0} S_n = } + \ 
   \sum_{l=0}^n \epsilon^{2l-2n} \sum_{k=0}^n (-1)^k
   \binom{n}{k}  b_{k,l} \ -2 \ 
   \sum_{l=0}^{n-1} \epsilon^{2l-2n+2} \sum_{k=0}^n (-1)^k
   \binom{n}{k} k \,\, b_{k,l}.
\label{sn1}
\end{equation}
When the well known result
\[
   \sum_{k=0}^n (-1)^k \binom{n}{k}
   k^l = (-1)^n \, n! \,\, \delta_{n,l} \x{3} {\rm for\ all\ } n,l \in\bbbn_0,\ 
   l \leq n {\rm \ and\ where\ } 0^0 \equiv 1
\]
is used, one can see that \req{sn1} can be written as
\begin{equation}
   \lim_{\epsilon\rightarrow0} S_n = \lim_{\epsilon\rightarrow0} \sum_{l=0}^n (-1)^l \,
   \frac{\epsilon^{2l-2n}}{l!} \, (-1)^n \, n! \,\, \delta_{n,l} +
   \sum_{l=0}^{n-1} (-1)^{l+1} \, \frac{2\epsilon^{2l-2n+2}}{l!} \, (-1)^n \,
   n! \,\, \delta_{n,l+1} =
   1 + \frac{2n!}{(n-1)!} = 2n + 1.
\label{sn2}
\end{equation}
The last two summations in \req{sn1} evaluate to zero because $b_{k,l}$ only
contains powers of $k$ up to $k^{l-1}$ or less. When \req{sn2} is substituted in
\req{gen:inf0} one finds the following result for the limit
$\beta\rightarrow\infty$
\begin{equation}
   \sum_{n=0}^\infty (-1)^n\,(2n+1)\,\frac{c_{2n}\p\,\gamma^{2n}\,\ln^n2}
   {n!} = 0.
\label{gen:inf}
\end{equation}
Which can be used to determine the limiting value of the conversion factor for
any arbitrary surface brightness distribution.  In the context of this
section, an arbitrary surface brightness profile can mean an arbitrary
observed profile, i.e. an intrinsic profile convolved with an arbitrary beam
profile. To indicate this difference with previous definitions, primes have
been added to the radial moments.  When \req{convfac} is used, this expression
can be rewritten into an expression which yields the \fwhm\ of the observed
profile directly:
\begin{equation}
   \sum_{n=0}^\infty (-1)^n \, \frac{2n+1}{n!} \  c_{2n}\p \, \left(
   \frac{4\ln2}{\Phi^2} \right)^n = 0.
\label{gen:inf2}
\end{equation}
This implicit equation uses only the radial moments of the observed profile
and can be solved easily using a Newton-Raphson scheme.

\section{Determining diameters using second moments}
\label{sec:mom}

It is well known that the \fwhm\ of a gaussian is related to the second moment
of a gaussian profile through:
\begin{equation}
  \Phi_2 = 2 \sqrt{\frac{c_2\,\ln2}{c_0}}.
\label{sec:def}
\end{equation}
This formula is widely used to calculate the \fwhm\ of an arbitrary profile.
In general however, the result will not be identical to the \fwhm\
derived from a gaussian fit. To distinguish the two values a subscript $2$
has been used. Since the definition of the \fwhm\ is not identical, also
the value for the conversion factor will be different.
One can define (using \reqb{phi1})
\begin{equation}
  \Phd = \sqrt{\Phi_2^2 - \Phb^2(p)} \x{3} \Rightarrow \x{3}
  \gamma_2(\beta) = \frac{2}{\Phd} = \left[ \ln2 \left( \frac{c_2\p}{c_0\p}
  - \frac{1}{p} \right) \right]^{-\frac{1}{2}}.
\label{gam:def2}
\end{equation}
To distinguish between the values of the radial moments for the unconvolved and
the convolved profile, a prime has been used in the latter case.
Now an expression for the radial moments of the convolved
profile will be derived.
Substituting \req{conv:func2} in \req{std:moment} one finds
\begin{equation}
  c_{2n}\p = 2\pi \, a\sum_{k=0}^\infty c_{2k}\,p^k
  \sum_{l=0}^k \,(-1)^{k-l}\,\frac{p^l}{(k-l)!\,l!^2}
  \int_0^\infty r^{2l+2n+1}\,\me^{-pr^2} \md r.
\label{lem:int3}
\end{equation}
Here the order of summation and integration has been reversed; in
Lemma~\ref{l:sunc3} it is proven that this is allowed provided condition~(2b)
is fulfilled.
The integral can evaluated easily using the substitution $s = r^2$ and yields
\[
  c_{2n}\p = \pi a\sum_{k=0}^\infty c_{2k}\,p^k \sum_{l=0}^k \,
  (-1)^{k-l}\,\frac{p^l}{(k-l)!\,l!^2} \frac{(l+n)!}{p^{l+n+1}} =
  \pi \frac{a}{p} \sum_{k=0}^\infty c_{2k}\,p^{k-n} \sum_{l=0}^k \,
  (-1)^{k-l}\,\frac{(l+n)!}{(k-l)!\,l!^2}.
\]
The fact that the gaussian representing the beam should be normalized
to 1 implies $a = p/\pi$. If $a_{n,i}$ is defined such that
\[
   \sum_{i=0}^n a_{n,i} \, l^i \equiv
   (l+n)\ldots(l+2)(l+1) = \frac{(l+n)!}{l!} \x{3}
   \hbox{for all } n,l\in\bbbn_0 \hbox{,\ and\ where\ } 0^0 \equiv 1
   \x{3} \Rightarrow
\]
\[
   a_{n,0} = n!,\x{3} a_{n,n} = 1,\x{3} a_{n,i} = n a_{n-1,i} + a_{n-1,i-1}\x{3}
   \hbox{for all }n\in\bbbn_0,\x{1} 1 \leq i \leq n-1
\]
one can write
\[
  c_{2n}\p = \sum_{k=0}^\infty c_{2k}\,\frac{p^{k-n}}{k!} \sum_{l=0}^k \,
  (-1)^{k-l}\,\frac{k!}{(k-l)!\,l!} \sum_{i=0}^n a_{n,i}\, l^i =
  \sum_{k=0}^\infty \frac{c_{2k}}{p^{n-k}}\,
  \sum_{i=0}^n (-1)^k\,\frac{a_{n,i}}{k!} \sum_{l=0}^k\,(-1)^l\,
  \binom{k}{l}\, l^i.
\]
The innermost summation is well known and yields non-zero results only
when $i \geq k$. Since $i \leq n$, this also implies $k \leq n$. Hence one
can write
\[
  c_{2n}\p = \sum_{k=0}^n \frac{c_{2k}}{p^{n-k}}\,
  \sum_{i=k}^n (-1)^k\,\frac{a_{n,i}}{k!} \sum_{l=0}^k\,(-1)^l\,
  \binom{k}{l}\, l^i \equiv
  \sum_{k=0}^n b_{n,k} \frac{c_{2k}}{p^{n-k}}.
\]
The following result will be postulated for the coefficients $b_{n,k}$
\[
  b_{n,k} = \sum_{i=k}^n (-1)^k\,\frac{a_{n,i}}{k!} \sum_{l=0}^k\,(-1)^l\,
  \binom{k}{l}\, l^i =
  \frac{n!^2}{(n-k)!\,k!^2}\x{3} \hbox{for all } n,k\in\bbbn_0,\x{1} k\leq n.
\]
This expression has been checked numerically as was found to be correct
for all $n,k \leq 19$. This makes it very plausible that it is correct
for all $n$ and $k$. Substituting this result yields the following expression
\begin{equation}
   c_{2n}\p = \sum_{k=0}^n \frac{n!^2}{(n-k)!\,k!^2} \, \frac{c_{2k}}{p^{n-k}}
   \x{3} \Rightarrow \x{3}
   c_0\p = c_0, \x{3} c_2\p = \frac{c_0}{p} + c_2, \x{3}
   c_4\p = \frac{2c_0}{p^2} + \frac{4c_2}{p} + c_4, \x{3}
   \cdots
\label{impl:expl}
\end{equation}
Substituting \req{impl:expl} in \req{gam:def2}, one finds
\begin{equation}
  \gamma_2(\beta) = \left[ \ln2 \left( \frac{c_2\p}{c_0\p}
  - \frac{1}{p} \right) \right]^{-\frac{1}{2}} =
  \left[ \ln2 \left( \frac{1}{p} + \frac{c_2}{c_0}
  - \frac{1}{p} \right) \right]^{-\frac{1}{2}} = \sqrt{\frac{c_0}{c_2\,\ln2}}
  = \gamma(0).
\label{gamma:sec}
\end{equation}
Thus it has been proven that the conversion factor for second
moment deconvolution is independent of beam size and equal to the
conversion factor for gaussian deconvolution in the limit for infinitely
large beams.

\section{Conclusions and future work}
\label{concl:iv}

In this work conversion factors have been determined to convert the
deconvolved \fwhm\ of a partially resolved nebula to its true diameter.  This
conversion factor depends on the \fwhm\ of the beam and the intrinsic surface
brightness distribution of the source.  All work in this paper has been
restricted to circularly symmetric surface brightness distributions and beams.
The following results were obtained.
\begin{enumerate}
\item
An implicit equation has been derived which can be used to determine
the conversion factor given the intrinsic surface brightness distribution,
the measured \fwhm\ and the beam size.
\item
From this implicit equation, various explicit expressions have been derived,
which give the conversion factor in cases where the beam size
is larger than the source.
\item
Finally the implicit equation is used to construct an alternative algorithm
for determining the \fwhm\ of an arbitrary observed surface brightness
distribution.
\item
The \fwhm\ derived with gaussian deconvolution is in general not equal to
the \fwhm\ derived with second moment deconvolution. Hence the
conversion factors will also be different for both methods.  The use of second
moment deconvolution is studied for the first time
in this paper and it is found that the conversion factor is {\em independent}
of the beam size in this case.  In the limit for infinitely large
beam sizes, the values of the conversion factors for both methods are
equal.
\end{enumerate}

\noindent
In a forthcoming publication the application of this theory to actual
observations will be discussed (van Hoof 1999). Particular attention
will be given to the limitations of the gaussian and second moment
deconvolution method. Also a new method for deconvolving angular diameters
will be presented.

\section*{Acknowledgments}
The author would like to thank G.C. Van de Steene for inspiring this research.
The author was supported by NFRA grant
782--372--033 during his stay in Groningen, and is currently supported by the
NSF through grant no.\ AST 96--17083.

\appendix

\section{Additional proofs}
\label{notes}

In this section additional proofs for the existence and convergence of certain
integrals and series will be presented.

\begin{lem}
\label{l:cn}
When $c_n$ exists, then also all $c_i$ exist with $i < n$.
\end{lem}
\begin{pf}
One can write
\[ c_i = 2\pi \! \int_0^\infty f(r)\,r^{i+1} \md r =
    2\pi \! \int_0^1 f(r)\,r^{i+1} \md r + 2\pi \! \int_1^\infty
    f(r)\,r^{i+1} \md r \leq \frac{2\pi F}{i+2} + 2\pi \! \int_1^\infty
    f(r)\,r^{n+1} \md r \leq \]
\[  \phantom{c_i} \leq \pi F + c_n - 2\pi\!\int_0^1 f(r)\,r^{n+1} \md r
    \leq \pi F + c_n. \]
Since $f(r)r^{n+1}$ is positive everywhere.
Hence the integral $c_i$ exists for all $i < n$.
\end{pf}

\begin{lem}
\label{cn:lim}
For $f(r)$ for which conditions~(1) and (2c) are
fulfilled, condition~(2b) is also fulfilled.
\end{lem}
\begin{pf}
One can write
\[
   c_{2n} = 2\pi \! \int_0^\infty f(r)\,r^{2n+1} \md r =
   2\pi \! \int_0^{R_o} f(r)\,r^{2n+1} \md r \leq
   2\pi\,F \! \int_0^{R_o} r^{2n+1} \md r = \frac{\pi\,F}{n+1} R_o^{2n+2}.
\]
This proves that all $c_{2n}$ with $n\in\bbbn_0$ exist. Now one can write
\[
   \lim_{n\rightarrow\infty} c_{2n}^{1/n} \leq \left( \frac{\pi\,F}{n+1}
   R_o^{2n+2} \right)^{1/n} = \lim_{n\rightarrow\infty} \left(
   \frac{\pi \, F \, R_o^2}{n+1} \right)^{1/n} \!\! R_o^2 = R_o^2
   \x{3} \Rightarrow \x{3} \lim_{n\rightarrow\infty} c_{n}^{1/n} = R_0
   = o(n^\frac{1}{2})
\]
and thus condition~(2b) is fulfilled.
\end{pf}

\begin{lem}
\label{l:exi}
The integral $I_{n,k,l}$ exists for all $n,k\in\bbbn_0$
and all $l\in\bbbn$ provided condition~(1) is fulfilled.
The integral $I_{n,k,0}$ exists for all
$n\in\bbbn_0$ and all $k\in\bbbn$, provided that $c_{2n}$ exists and
condition~(1) is fulfilled.
\end{lem}
\begin{pf}
To prove the first part one can simply write
\[ I_{n,k,l} = \int_0^\infty f^k(r) g^l(r;a,p)\,r^{2n+1} \md r \leq
  a^l\,F^k \int_0^\infty \me^{-lpr^2} r^{2n+1} \md r =
  \frac{a^l\,F^k\,n!}{2\,(lp)^{n+1}}. \]
To prove the second part one can write
\[ I_{n,k,0} = \int_0^\infty f^k(r)\,r^{2n+1} \md r \leq
   F^{k-1} \int_0^\infty f(r)\,r^{2n+1} \md r = 
   \frac{F^{k-1}}{2\pi}\,c_{2n}, \]
and hence the integral exists if $c_{2n}$ exists.
\end{pf}

\begin{lem}
\label{l:iunc}
The integral $I_{n,k,l}$ is uniformly convergent for all
$a,p\in(0,\infty)$, for any $n,k\in\bbbn_0$
and any $l\in\bbbn$, provided condition~(1) is fulfilled.
\end{lem}
\begin{pf}
Lemma~\ref{l:exi} proves that
the integral $I_{n,k,l}$ exists. Now one can use
\[ I^\prime_{n,k,l} = \int_X^Y f^k(r) g^l(r;a,p) r^{2n+1} \md r
  \leq a^lF^k \int_X^Y \me^{-lpr^2} r^{2n+1} \md r. \]
If the substitution $s = r^2$ is used, one finds
\[ I^\prime_{n,k,l} \leq \frac{a^lF^k}{2} \int_{X^2}^{Y^2} \me^{-lps}
  s^n \md s = \frac{a^lF^k}{2(lp)^{n+1}} \sum_{i=0}^n \frac{n!}{(n-i)!} \left[
  (lpX^2)^{n-i}\,\me^{-lpX^2} - (lpY^2)^{n-i}\,\me^{-lpY^2} \right]. \]
Since $lpY^2 > 0$, one may write
\[ I^\prime_{n,k,l} \leq \frac{a^lF^k}{2(lp)^{n+1}} \sum_{i=0}^n \frac{n!}
  {(n-i)!}(lpX^2)^{n-i}\,\me^{-lpX^2}. \]
Since $n!/(n-i)! \leq n!$ and $(lpX^2)^{n-i} \leq (lpX^2)^n$
for $X \geq (lp)^{-\frac{1}{2}}$, one may write
\[ I^\prime_{n,k,l} \leq \frac{a^lF^k}{2(lp)^{n+1}}\,(n+1)\,n!\,(lpX^2)^n\,
  \me^{-lpX^2} = \frac{(n+1)!\,a^lF^k}{2lp} X^{2n}\,\me^{-lpX^2}. \]
It is well known that $\lim_{X\rightarrow\infty} X^{2n} \me^{-lpX^2} = 0$
for any $n\in\bbbn_0$, $l\in\bbbn$ and $p\in(0,\infty)$. Hence one can
always find a finite solution $X \geq (lp)^{-\frac{1}{2}}$ for the inequality
\[ |I^\prime_{n,k,l}| < \epsilon \x{3} \Rightarrow \x{3}
   X^{2n} \me^{-lpX^2} < \frac{2lp\,\epsilon}{(n+1)!\,a^l\,F^k} \]
for all $\epsilon>0$, $n,k\in\bbbn_0$, $l\in\bbbn$ and $a,p\in(0,\infty)$.
This proves that $I_{n,k,l}$ is uniformly convergent.
\end{pf}

\begin{lem}
\label{l:sunc1}
In the expression given in \req{lem:int} the order of
summation and integration may be reversed for arbitrary values of $p$ and $r$,
provided that conditions~(1) and
(2a) are fulfilled.
\end{lem}
\begin{pf}
Since the terms in \req{lem:int} have alternating signs,
the following partial sum will be considered
\[
  I_{N,R} = \sum_{n=0}^N \sum_{i=0}^n \int_0^R \int_0^{2\pi} \left|
  \, a \,\me^{-p(x^2+y^2)} f(r\p)\,\me^{-pr\p^2}
  \,(2p)^n \frac{x^{n-i}y^i}{(n-i)!\,i!}\,r\p^n\cos^{n-i}\!\varphi\p\,
  \sin^i\!\varphi\p \, r\p \right| \md\varphi\p \md r\p.
\]
After changing $x$, $y$ to radial coordinates, one can write
\[
  I_{N,R} \leq \sum_{n=0}^N \sum_{i=0}^n \int_0^R \int_0^{2\pi}
  a\,(2p)^n \frac{r^n}{(n-i)!\,i!}\,f(r\p)\,r\p^{n+1} \md\varphi\p \md r\p =
  \sum_{n=0}^N \sum_{i=0}^n \int_0^R
  2\pi a \,(2p)^n \frac{r^n}{n!} \binom{n}{i} f(r\p)\,r\p^{n+1} \md r\p =
\]
\[
  \phantom{I_{N,R}} = \sum_{n=0}^N \sum_{i=0}^n 2\pi a \,
  \frac{(2pr)^n}{n!} \binom{n}{i}
  \int_0^R f(r\p)\,r\p^{n+1} \md r\p \leq
  \sum_{n=0}^N \sum_{i=0}^n 2\pi a \, \frac{(2pr)^n}{n!}
  \binom{n}{i} c_n,
\]
where for the last step condition~(1) has been used.
Since it is assumed that $c_{2n}$ exists for all $n\in\bbbn_0$, this implies
that $c_n$ exists for all $n\in\bbbn_0$ (Lemma~\ref{l:cn}). Hence $I_{N,R}$
exists for arbitrary $N$ and $R$. This also proves that
$\lim_{R\rightarrow\infty} I_{N,R}$ exists. For arbitrary $R$ the following
is true
\[
  I_{N,R} \leq \sum_{n=0}^N 2\pi a \, \frac{(2pr)^n}{n!}\, c_n
  \sum_{i=0}^n \binom{n}{i} =
  \sum_{n=0}^N 2\pi a \, \frac{(2pr)^n}{n!}\, c_n \, 2^n =
  \sum_{n=0}^N 2\pi a \, \frac{(4pr)^n}{n!}\, c_n \equiv 
  \sum_{n=0}^N a_n.
\]
In order for $\lim_{N\rightarrow\infty} I_{N,R}$ to exist, the following must
be true
\[
  \lim_{n\rightarrow\infty} a_n^{1/n} = \lim_{n\rightarrow\infty} \left[ 2\pi
  a \frac{(4pr)^n}{n!}\, c_n \right]^{1/n} < \lim_{n\rightarrow\infty} \left[
  2\pi a \frac{(4p\,\me\,r)^n}{n^n} c_n \right]^{1/n} =
  \lim_{n\rightarrow\infty} (2\pi a)^{1/n} \frac{4p\,\me\,r}{n} c_n^{1/n} =
  \lim_{n\rightarrow\infty} \frac{4p\,\me\,r}{n} c_n^{1/n} < 1.
\]
Here the well known result $n! > \me^{-n} n^n$ for all $n > 0$ has been used.
In order for the last inequality to be true for arbitrary values of $p$ and
$r$, the constraint $c_n^{1/n} = o(n)\,(n\rightarrow\infty)$ has to be
imposed.  This proves that $\lim_{N\rightarrow\infty} I_{N,R}$ exists. Since
the argument given above is also valid in the limit $R\rightarrow\infty$, this
also proves that $\lim_{N,R\rightarrow\infty} I_{N,R}$ exists. This completes
the proof that the order of integration and summation in \req{lem:int} may be
reversed.
\end{pf}

\begin{lem}
\label{l:cnp}
The series given in \req{conv:func} is absolutely and
uniformly convergent for all $r\in[0,\infty)$ and $p\in[0,\infty)$, provided
that conditions~(1) and (2a) are fulfilled.
\end{lem}
\begin{pf}
First it will be proven that these
conditions imply $[c_n(p)]^{1/n} = o(n)\,(n\rightarrow\infty)$ for all
$p\in[\,0,\infty\,)$.  Since $f(r) \geq 0$ everywhere, one can write
\[ c_{2n}(p) = 2\pi \! \int_0^\infty f(r) \, r^{2n+1} \me^{-pr^2} \md r
   \leq 2\pi\!\int_0^\infty f(r)\, r^{2n+1} \md r = c_{2n}. \]
This proves that $c_{2n}(p)$ exists for all $n\in\bbbn_0$ and all
$p\in[\,0,\infty\,)$. It also implies
\[ [c_n(p)]^{1/n} \leq c_n^{1/n} = o(n)\,(n\rightarrow\infty), \]
for all $p\in[\,0,\infty\,)$, which completes the first step. The summation
in \req{conv:func} is a Taylor series, which will have an infinite radius of
convergence when the following condition is fulfilled
\[
  \lim_{n\rightarrow\infty} \left[ \frac{c_{2n}(p) p^{2n}}
  {n!^2} \right]^{1/(2n)} < \lim_{n\rightarrow\infty} \left[
  \frac{c_{2n}(p) (p\,\me)^{2n}}{n^{2n}} \right]^{1/(2n)} =
  \lim_{n\rightarrow\infty} \frac{p\,\me}{n} \, [c_{2n}(p)]^{1/(2n)} = 0.
\]
Here the well known result $n! > \me^{-n} n^n$ for all $n > 0$ has been used.
For the last condition to hold for any arbitrary value of $p$, it must be true
that $[c_{2n}(p)]^{1/(2n)} =
o(n)\,(n\rightarrow\infty)$ or $[c_n(p)]^{1/n} = o(n)\,(n\rightarrow\infty)$,
which has been proven to be true. Any Taylor series is absolutely and uniformly
convergent within its radius of convergence. Since the gaussian in
\req{conv:func} is bounded for all $p,r\in[0,\infty)$, this also proves
that the expression in  \req{conv:func} is absolutely and uniformly
convergent for all $p,r\in[0,\infty)$.
\end{pf}

\begin{lem}
\label{l:sunc2}
In the expression given in \req{lem:int2} the order of
summation and integration may be reversed for arbitrary values of $p$,
provided that conditions~(1) and
(2b) are fulfilled.
\end{lem}
\begin{pf}
Since the terms in \req{lem:int2} have alternating signs,
the following partial sum will be considered
\[
  I_{K,R} = \sum_{k=0}^K \int_0^R \left| \, 2\pi\,(-1)^k\,\frac{p^k}{k!}\,
  f(r)\,r^{n+2k+1} \right| \md r = \sum_{k=0}^K 2\pi\,\frac{p^k}{k!}
  \int_0^R f(r)\,r^{n+2k+1} \md r \leq
  \sum_{k=0}^K 2\pi\,
  \frac{p^k}{k!} \, c_{n+2k} \equiv \sum_{k=0}^K a_k,
\]
where for the last inequality condition~(1) has been used.  Since it is
assumed that $c_{2n}$ exists for all $n\in\bbbn_0$, this implies that $c_n$
exists for all $n\in\bbbn_0$ (Lemma~\ref{l:cn}).  From this it is clear that
$I_{K,R}$ exists for arbitrary values of $K$ and $R$.  It is also clear that
$\lim_{R\rightarrow\infty} I_{K,R}$ exists for any value of $K$. In order for
$\lim_{K\rightarrow\infty} I_{K,R}$ to exist, the following must be true
\[
  \lim_{k\rightarrow\infty} a_k^{1/k} = \lim_{k\rightarrow\infty} \left(
  2\pi\, \frac{p^k}{k!} \, c_{n+2k} \right)^{1/k} < \lim_{k\rightarrow\infty}
  \left[ 2\pi\,\frac{(p\,\me)^k}{k^k} \, c_{n+2k} \right]^{1/k} =
  \lim_{k\rightarrow\infty} (2\pi)^{1/k}\,\frac{p\,\me}{k}\, c_{n+2k}^{1/k} =
  \lim_{k\rightarrow\infty} \frac{p\,\me}{k}\, c_{2k}^{1/k} < 1.
\]
Here the well known result $k! > \me^{-k} k^k$ for all $k > 0$ has been used.
In order for the last inequality to be true for arbitrary values of $p$, the
constraint $c_{2k}^{1/k} = o(k)\,(k\rightarrow\infty)$ has to be
imposed. Using the substitution $n = 2k$, this can be written as
$c_n^{1/n} = o(n^\frac{1}{2})\,(n\rightarrow\infty)$.  This completes the
proof that $\lim_{K\rightarrow\infty} I_{K,R}$ exists.  Since the argument
given above is also valid in the limit $R\rightarrow\infty$ this also proves
that $\lim_{K,R\rightarrow\infty} I_{K,R}$ exists.  This completes the proof
that the order of integration and summation in \req{lem:int2} may be reversed.
\end{pf}

\begin{lem}
\label{l:sunc5}
The series given in \req{tayl:moment} is absolutely and
uniformly convergent for all $p\in[0,\infty)$, provided condition~(2b) is
fulfilled.
\end{lem}
\begin{pf}
The summation in \req{tayl:moment} is a Taylor series in $p$, which
will have an infinite radius of convergence when the following criterion is
fulfilled
\[
  \lim_{k\rightarrow\infty} \left( \frac{c_{n+2k}}{k!}
  \right)^{1/k} < \lim_{k\rightarrow\infty} \left( \frac{\me^k}{k^k} \,
  c_{n+2k} \right)^{1/k} = \lim_{k\rightarrow\infty}
  \frac{\me}{k} \, c_{n+2k}^{1/k} = \lim_{k\rightarrow\infty}
  \frac{\me}{k}\,c_{2k}^{1/k}  = 0.
\]
Here the well known result $k! > \me^{-k} k^k$ for all $k > 0$ has been used.
Condition~(2b) assures that this is true, which can be seen by
making the transformation $n=2k$. This proves that the series has an infinite
radius of convergence. Since any Taylor series is absolutely and uniformly
convergent within its radius of convergence, this completes the proof.
\end{pf}

\begin{lem}
\label{l:sunc4}
In the expression given in \req{lem:int4} the order of
summation and integration may be reversed for arbitrary values of $p$,
provided that condition~(2b) is fulfilled.
\end{lem}
\begin{pf}
Since the terms in \req{lem:int4} don't always have the
same sign, the following partial sum will be considered
\[
   I_{N,S} = \sum_{n=0}^N\,\int_0^S \left|\, (1 - 2s) \, \me^{-\zeta s} 
   c_{2n}p^n \sum_{k=0}^n (-1)^{n-k}\frac{(\zeta - 1)^k}{(n-k)!\,k!^2} s^k
   \right| \md s \leq
   \sum_{n=0}^N c_{2n}\,p^n \sum_{k=0}^n\,\frac{(\zeta - 1)^k}
   {(n-k)!\,k!^2} \int_0^S (2s + 1)\, s^k \, \me^{-\zeta s} \md s \leq
\]
\[
   \phantom{I_{N,S}} \leq
   \sum_{n=0}^N\,c_{2n}\,p^n \sum_{k=0}^n\,\frac{(\zeta - 1)^k}
   {(n-k)!\,k!^2} \left[ 2 \frac{(k+1)!}{\zeta^{k+2}} + \frac{k!}{\zeta^{k+1}}
   \right] \leq
   \sum_{n=0}^N\,\frac{c_{2n}\,p^n}{\zeta^2} \sum_{k=0}^n\,\frac{\zeta+2k+2}
   {(n-k)!\,k!} \leq
\]
\[
   \phantom{I_{N,S}} \leq
   \sum_{n=0}^N\,\frac{c_{2n}\,p^n}{n!}\frac{(\zeta+2n+2)}{\zeta^2}
   \sum_{k=0}^n\,\binom{n}{k} =
   \sum_{n=0}^N\,\frac{c_{2n}\,(2p)^n}{n!}\frac{(\zeta+2n+2)}{\zeta^2} \equiv
   \sum_{n=0}^N a_n.
\]
Here the fact has been used that $\zeta > 1$ (see \reqb{zeta:beta}) and hence
$(\zeta-1)^k < \zeta^k$. Since it is assumed that $c_{2n}$ exists for all
$n\in\bbbn_0$, it is clear that $I_{N,S}$ exists for arbitrary values of $N$
and $S$. It is also clear that $\lim_{S\rightarrow\infty} I_{N,S}$ exists. In
order for $\lim_{N\rightarrow\infty} I_{N,S}$ to exist, the following must be
true
\[
   \lim_{n\rightarrow\infty} a_n^{1/n} = \lim_{n\rightarrow\infty}
   \left[ \frac{c_{2n}\,(2p)^n}{n!}\frac{(\zeta+2n+2)}{\zeta^2} \right]^{1/n} <
   \lim_{n\rightarrow\infty} \left[ \frac{c_{2n}\,(2p\,\me)^n}{n^n}
   \frac{(\zeta+2n+2)}{\zeta^2} \right]^{1/n} = 
\]
\[
   \phantom{\lim_{n\rightarrow\infty} a_n^{1/n}} =
   \lim_{n\rightarrow\infty} \left[ c_{2n}\frac{(\zeta+2n+2)}{\zeta^2}
   \right]^{1/n} \frac{2p\,\me}{n} =
   \lim_{n\rightarrow\infty} c_{2n}^{1/n} \frac{2p\,\me}{n} < 1.
\]
Here the well known result $n! > \me^{-n} n^n$ for all $n > 0$ has been used.
In order for the last inequality to be true for arbitrary values of $p$, the
constraint $c_{2n}^{1/n} = o(n)\,(n\rightarrow\infty)$ has to be
imposed. Using the substitution $2n\rightarrow n$, this can be written as
$c_n^{1/n} = o(n^\frac{1}{2})\,(n\rightarrow\infty)$.  This completes the
proof that $\lim_{N\rightarrow\infty} I_{N,S}$ exists.  Since the argument
given above is also valid in the limit $S\rightarrow\infty$ this also proves
that $\lim_{N,S\rightarrow\infty} I_{N,S}$ exists.  This completes the proof
that the order of integration and summation in \req{lem:int4} may be reversed.
\end{pf}

\begin{lem}
\label{lem:akl}
The expression for $a_{k,l}$ given in \req{sn4} yields
a polynomial in $k$ of degree $l$ for all $l\in\bbbn_0$.
\end{lem}
\begin{pf}
From the definition given in \req{sn4} it is clear that the lemma
is correct for $l=0$. If it is assumed that the lemma is correct for all
$l^\prime < l$ then the last part of \req{sn4} assures that $a_{k,l}$ is equal
to the sum of $l$ polynomials of degree $l^\prime = l-1,\ l-2,\ \ldots$,
multiplied by
a linear term in $k$. This yields a polynomial of degree $\max(l^\prime+1) = l$.
Since the lemma is true for $l=0$, this proves by induction that the lemma
is true for all $l\in\bbbn_0$.

Now an expression for the highest order term of this polynomial will be
derived. It is clear that the highest order contribution
to $a_{k,l}$ only comes from the term with $m=1$ in \req{sn4}. If one defines
\[ a_{k,l} = \sum_{i=0}^l d_{l,i} k^i, \]
one can deduce the following expression for $d_{l,l}$ from \req{sn4}
\[
   d_{l,l} = -\frac{k}{l} d_{l-1,l-1} = \frac{k^2}{l(l-1)} d_{l-2,l-2} =
   \left[ \prod_{i=0}^l (-1)\frac{k}{i} \right] d_{0,0} = (-1)^l \frac{k^l}{l!}
   \x{3} \Rightarrow \x{3}
   a_{k,l} = (-1)^l \frac{k^l}{l!} + \sum_{i=0}^{l-1} d_{l,i} k^i \equiv
   (-1)^l \frac{k^l}{l!} + b_{k,l}.
\]
Here $b_{k,l}$ is the lower order remainder of $a_{k,l}$ which is a polynomial
of degree $l-1$ or less in $k$.
It can easily be checked that these expressions are also valid for $k=0$
and $l=0$ if one postulates $0^0 \equiv 1$ and $b_{k,0} \equiv 0$.
\end{pf}

\begin{lem}
\label{l:sunc3}
In the expression given in \req{lem:int3} the order of
summation and integration may be reversed for arbitrary values of $p$,
provided that condition~(2b) is fulfilled.
\end{lem}
\begin{pf}
Since the terms in \req{lem:int3} have alternating signs,
the following partial sum will be considered
\[
  I_{K,R} = \sum_{k=0}^K \int_0^R \left| \, 2\pi \, a c_{2k}p^k 
  \sum_{l=0}^k \,(-1)^{k-l}\,\frac{p^l}{(k-l)!\,l!^2} r^{2l+2n+1}\,
  \me^{-pr^2} \right| \md r = 
  \sum_{k=0}^K \,2\pi a\,c_{2k}\,p^k \sum_{l=0}^k \,\frac{p^l}{(k-l)!\,l!^2}
  \int_0^R r^{2l+2n+1}\,\me^{-pr^2} \md r \leq
\]
\[
  \phantom{I_{K,R}} \leq
  \sum_{k=0}^K \,\pi a\,c_{2k}\,p^k \sum_{l=0}^k \,\frac{p^l}{(k-l)!\,l!^2}
  \frac{(l+n)!}{p^{l+n+1}} =
  \sum_{k=0}^K \,\pi\frac{a}{p}\,c_{2k}\,p^{k-n} \sum_{l=0}^k\,\frac{(l+n)!}
  {(k-l)!\,l!^2} =
  \sum_{k=0}^K \,c_{2k}\,p^{k-n} \sum_{l=0}^k \,
  \frac{k!}{(k-l)!\,l!}\frac{(l+n)!}{l!\,k!} \leq
\]
\[
  \phantom{I_{K,R}} \leq
  \sum_{k=0}^K \,c_{2k}\,p^{k-n}
  \sum_{l=0}^k \binom{k}{l}
  \frac{(l+n)^n}{k!} \leq
  \sum_{k=0}^K \, c_{2k}\,p^{k-n} \frac{(k+n)^n}{k!} \sum_{l=0}^k
  \binom{k}{l}  =
  \sum_{k=0}^K \, c_{2k}\,p^{k-n} \frac{2^k\,(k+n)^n}{k!} \equiv
  \sum_{k=0}^K a_k.
\]
In this derivation it has been used that the gaussian used to convolve the
profile should be normalized to 1, this implies $a = p/\pi$.  Since it is
assumed that $c_{2k}$ exists for all $k\in\bbbn_0$, this proves that $I_{K,R}$
exists for arbitrary values of $K$ and $R$. It also proves that
$\lim_{R\rightarrow\infty} I_{K,R}$ exists for arbitrary values of $K$. In
order for $\lim_{K\rightarrow\infty} I_{K,R}$ to exist, the following must be
true
\[
   \lim_{k\rightarrow\infty} a_k^{1/k} = \lim_{k\rightarrow\infty}
   \left[ c_{2k}\,p^{k-n} \frac{2^k\,(k+n)^n}{k!} \right]^{1/k} =
   \lim_{k\rightarrow\infty} \left[ c_{2k} \left( \frac{k+n}{p} \right)^n
   \frac{(2p)^k}{k!} \right]^{1/k} <
   \lim_{k\rightarrow\infty}
   \left[ c_{2k} \left( \frac{k+n}{p} \right)^n \frac{(2p\,\me)^k}{k^k}
   \right]^{1/k} =
\]
\[
   \phantom{\lim_{k\rightarrow\infty} a_k^{1/k}} =
   \lim_{k\rightarrow\infty} c_{2k}^{1/k} \left( \frac{k+n}{p} \right)^{n/k}
   \frac{2p\,\me}{k} =
   \lim_{k\rightarrow\infty} c_{2k}^{1/k}\,\frac{2p\,\me}{k} < 1.
\]
Here the well known result $k! > \me^{-k} k^k$ for all $k > 0$ has been used.
In order for the last inequality to be true for arbitrary values of $p$, the
constraint $c_{2k}^{1/k} = o(k)\,(k\rightarrow\infty)$ has to be
imposed. Using the substitution $n = 2k$, this can be written as
$c_n^{1/n} = o(n^\frac{1}{2})\,(n\rightarrow\infty)$.  This completes the
proof that $\lim_{K\rightarrow\infty} I_{K,R}$ exists.  Since the argument
given above is also valid in the limit $R\rightarrow\infty$ this also proves
that $\lim_{K,R\rightarrow\infty} I_{K,R}$ exists.  This completes the proof
that the order of integration and summation in \req{lem:int3} may be reversed.
\end{pf}

\section{Symbols}
\label{symbols:iv}

The following definitions for the symbols have been used.

\mbtlist{$g(r;a,p)$}
\mbt{$a$}{Measure for the height of the gaussian.}
\mbt{$c_n$}{Radial moments of the surface brightness profile. Defined in 
    \req{std:moment}.}
\mbt{$c_n(p)$}{Generalized radial moments of the surface brightness profile.
    Defined in \req{gen:moment}.}
\mbt{$f(r)$}{The surface brightness profile of the nebula.}
\mbt{$F$}{Upper bound for the surface brightness profile.}
\mbt{$g(r;a,p)$}{Two-dimensional gaussian profile.}
\mbt{$j_\nu$}{The emissivity per unit frequency.}
\mbt{$L_n$}{Laguerre polynomial of degree $n$.}
\mbt{$p$}{Measure for the width of a gaussian. Related to the \fwhm\ through
    \req{phi1}.}
\mbt{$r$}{Polar coordinate on the sky. Subscripts have the following meaning:\\
    $i$ -- the inner radius of the nebula,\\
    $s$ -- the \str\ or outer radius of the nebula.}
\mbt{$x$,$y$}{Cartesian coordinates on the sky.}
\mbt{$\beta$}{Ratio of the deconvolved \fwhm\ diameter of the nebula
    to the \fwhm\ of the beam.}
\mbt{$\gamma$}{Factor to convert the deconvolved \fwhm\ to the
    true angular diameter. Subscripts have the following meaning:\\
    $<$none$>$ -- gaussian fits were used to determine the \fwhm\\
    $2$ -- second moments were used to determine the \fwhm\ (\reqb{gam:def2}).\\
    $f$ -- indicates the fitting function used to approximate $\gamma(\beta)$.}
\mbt{$\Gamma$}{The gamma function.}
\mbt{$\delta_{i,j}$}{Kronecker delta symbol.}
\mbt{$\zeta$}{Auxiliary variable, related to $\beta$ through \req{zeta:beta}.}
\mbt{$\dia$}{True angular diameter of the nebula ($\dia\equiv 2\rs$).}
\mbt{$\varphi$}{Polar coordinate on the sky.}
\mbt{$\Phi$}{The \fwhm\ diameter of a profile. Subscripts have the following
    meaning:\\
    $<$none$>$ -- indicates the \fwhm\ of a gaussian profile,\\
    $b$ -- indicates the \fwhm\ of the beam,\\
    $d$ -- indicates the deconvolved \fwhm\ of a profile (defined in
      \reqb{phi2} or \reqb{gam:def2}),\\
    $2$ -- indicates the \fwhm\ of a profile obtained using second
       moments (\reqb{sec:def}).}
\endmbtlist

\noindent
Other symbols have varying meanings as defined in the pertinent sections.

\end{document}